# An Alternative Approach to Data Acquisition Using Keyboard Emulation Technique

Shahrukh Khalid* and Adnan Ali Khan**
* Hamdard University, Karachi, Pakistan, ** University of Manchester, Manchester, United Kingdom

*Abstract*—A number of data acquisition systems depend on human interface to access computer for measuring, processing and analyzing data and to prepare it for presentation and storage. Data acquisition software is installed on the computer and all intended operations are performed manually. The data acquisition software requires user intervention for operations like selection of measurement setup, acquisition and storage of data to computer. The duty of users becomes laborious if the data acquisition process lasts for a long duration and requires continuous repetition of steps. An appropriate solution to overcome such problem is to replace the physical operator with a virtual user. This software generated simulated user sits at the data acquisition process through out and automate all the intended steps of data acquisition. This paper presents a new approach for data acquisition by using keyboard emulation technique. A keyboard emulation software is developed which runs beside the main data acquisition software and acts as a virtual user. All the operations which require user interface are performed through fully automated computer program. The developed software/system is executed in a real time environment and the functionality of the software is verified. In the end, potential application areas of the designed keyboard emulation software are explored.

## I. INTRODUCTION

New trends and ideas are continuously evolving in the area of Human Computer Interaction and consequently various data acquisition processes have been introduced based on application requirements. Keyboard is a longstanding interface device and still used as a fundamental input device for interaction with computer application and data acquisition softwares. Apart from keyboard, a very good account for alternative means of provision of text based input to software systems is provided in [1]. This includes mobile phone keypad, chord keyboard, pinch keyboard, Pen and tablet technique, Pen based Qwerty keyboard, Pen based disk keyboard and hand written character recognition. More easy to use speech based interface between man and machine is stressed in [2].

In many scenarios of data acquisition from test and measurement equipment, a continuous human computer interaction is required for performing various operations like measuring, processing, analyzing, presentation and storage of data. Physical user input is required from the keyboard for performing these operations. If the process is ongoing and extends to prolonged interval then the task of operator becomes very hectic and laborious as it requires continuous intervention. There are also other scenarios specially at hazardous places where a physical user cannot be present. Thus, in such cases data acquisition through physical user becomes impossible.

In order to solve such problems, a novel keyboard emulation based technique is used. In this method, a separate software is designed which runs beside the main data acquisition software and performs all the operations which are needed to be performed by a physical human user. This software can be termed as a virtual user as it generates all the necessary keyboard operations as and when required. Programming languages like Microsoft® Visual Basic and Visual C offer easy approach in designing of such softwares. These languages provide developers Application Programming Interfaces (APIs) which can be declared and invoked in the main program for generating keyboard events. An API can be contrasted with a graphical user interface or a command interface to an operating system or a program as shown in Figure 1.

APIs are called for generating the desired key. A sequence of data operation which is required for a process can be first written in terms of Algorithm. This algorithm is then converted into a program which runs besides the main data acquisition software and operates it like a human user for the entire duration of the process.

In this paper we have worked out a technique to apply Keyboard Emulation technique for automatically running the data acquisition softwares as per their defined data acquisition cycles. In Section II existing use of Keyboard Emulation technique in everyday life is discussed. Section III presents the software specifications and discusses about the various implementation issues. Section IV gives the use case of software implementation through a flow chart. Section V presents the sought application areas of data acquisition where this technique can be employed for increasing ease of use and process efficiency. Section VI gives the concluding remarks.

## II. EXISTING USE OF KEYBOARD EMULATION TECHNOLOGY

At present, keyboard emulation technology is used for variety of applications most popular of which are discussed in the proceeding section.

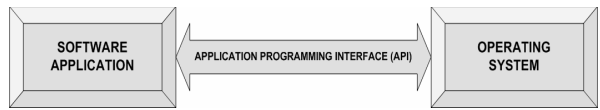

Figure 1. The application talks to the OS via API which defines parameters passed between them

### A. Helping Disabled Persons

Keyboard Emulation is used in helping disabled person in their various needs in using computer applications. Special hardware interfaces in the form of input devices which resembles like joysticks are used for acquiring input from disabled persons. The output of these interfaces is gathered in the special software application which converts these outputs to regular keyboard scan codes. Due to its role in provision of special needs for disabled persons, [3] refers such systems as electronic keyboard, surrogate keyboard or visual keyboard. Moreover [4] has proposed approach for designing assisted system as per the individual users capabilities along with preference and describes the usage of keyboard emulation. In [5] keyboard emulation through acoustic gestures is proposed. Figure 2 illustrates the idea that acoustic gestures can be recognized as a valid keyboard input through pattern matching algorithm.

### B. Keyboard Wedge Interface

Barcode scanners, weighing scales and other like devices are interfaced with computers through Keyboard. These devices generate Keyboard scan codes and its data is accepted into the software application for further processing. The application software is designed to accept only Keyboard scan code therefore it will be problematic to interface a new device which follows RS 232 based protocol. In order to resolve such bottlenecks, hardware and software based virtual keyboard implementation referred as Keyboard Wedge interface exists which converts RS 232 data to Keyboard scan code so that the software application can accept it. Hardware based emulator interface with the RS232 based device and converts RS 232 data to Keyboard scan code to be gathered through the Keyboard port of computer as shown in Figure 3. Whereas software based implementation accepts the RS 232 data of device attached at the RS 232 port of computer and converts it to Keyboard scan codes for the destined software application as shown in Figure 4. A practical implementation of hardware and software based keyboard wedges are available at [6]-[7]. Keyboard Emulation Hardware implemented by Altek instruments inc is shown in Figure 5.

### C. Game Industry

Another use of keyboard emulation based techniques exists in game industries in which keyboard emulators are designed for joysticks. The games are written to be played from keyboard but joysticks are preferred choice for users and give ease of use in game play. Therefore instead of programming for joysticks a keyboard emulator is used which programs the buttons of joystick and whenever any key is pressed from the joystick it converts it with keyboard scancode and OS considers it as a valid keyboard input. Figure 6 shows the picture of a keyboard emulator for joysticks.

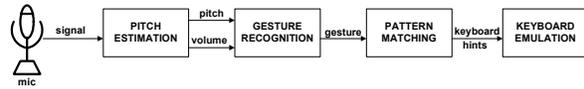

Figure 2. Keyboard emulation through acoustic gestures [5]

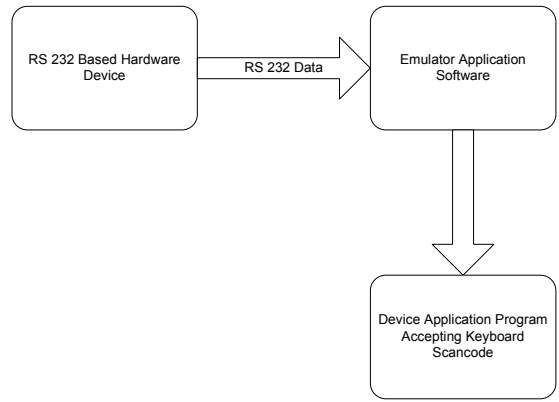

Figure 3: Software based Keyboard Emulator data flow

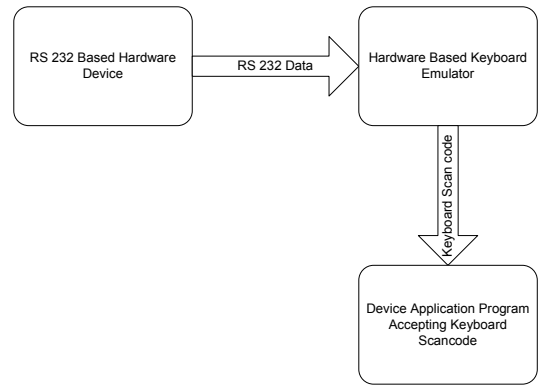

Figure 4  Hardware based Keyboard Emulator data flow

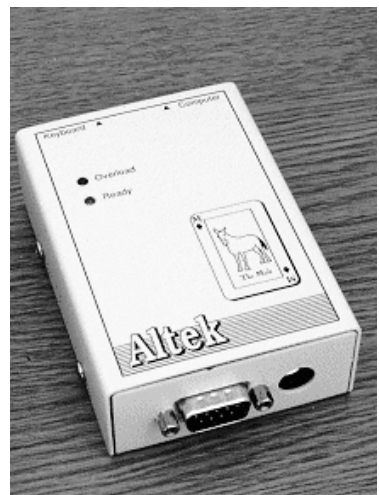

Figure 5: Keyboard Emulator for Barcode scanners from Altek Instruments Inc [6]

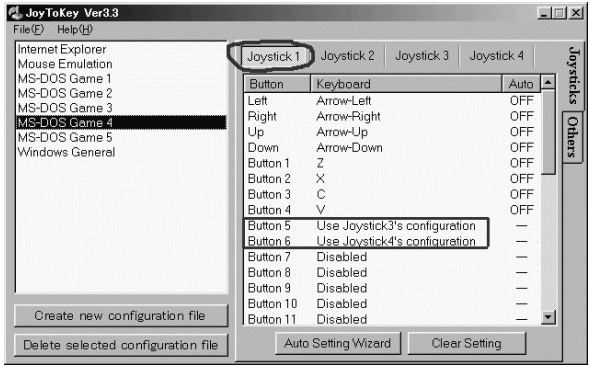

Figure 6: Snapshot of Keyboard Emulator for Joystick JoyToKey Ver 3.3 [8]

TABLE I. US KYEBOARD VIRTUAL KEYCODES

| Symbolic constant name | Value (hexadecimal) | Keyboard (or mouse) equivalent |
|---|---|---|
| VK_RETURN | 0D | ENTER key |
| VK_SHIFT | 10 | SHIFT key |
| VK_ESCAPE | 1B | ESC key |
| VK_SPACE | 20 | SPACEBAR |
| VK_PRIOR | 21 | PAGE UP key |
| VK_NEXT | 22 | PAGE DOWN key |
| VK_END | 23 | END key |
| VK_LEFT | 25 | LEFT ARROW key |
| VK_UP | 26 | UP ARROW key |
| VK_0 | 30 | 0 key |
| VK_1 | 31 | 1 key |
| VK_2 | 32 | 2 key |
| VK_4 | 34 | 4 key |
| VK_5 | 35 | 5 key |
| VK_6 | 36 | 6 key |
| VK_7 | 37 | 7 key |
| VK_8 | 38 | 8 key |
| VK_9 | 39 | 9 key |
| VK_A | 41 | A key |
| VK_B | 42 | B key |
| VK_D | 44 | D key |

## III. SOFTWARE SPECIFICATIONS

In this work, we have designed an emulator that follows the Keyboard Scan Code Specification Revision 1.3a published in March 2000 by Microsoft Corp [9]. It implements Scan Code Set 2 emulation. Scan Code Set 2 is the current standard for PC compatible computers. It is compatible with all current operating systems including DOS, Windows and Unix style Operating Systems like Linux and FreeBSD.

Windows defines special constants for each key the user can press. These constants are referred as virtual Key codes. These constants can then be used to refer to the keystroke when using Windows API calls. Virtual keys mainly consist of actual keyboard keys, but also include "virtual" elements such as the three mouse buttons. Table I gives some US Keyboard Virtual Key Codes.

## IV. SOFTWARE USE CASE

Design of Keyboard Emulation Software is based on the win32 API named as keyb_event. This API is used in the keyboard Emulation software for sending keyboard keys to the main data acquisition software. All the keys of a typical keyboard are declared in terms of their equivalent hexadecimal number in the program. The keybd_event function synthesizes a keystroke by sending the equivalent hexadecimal number. The keyboard drivers interrupt handler calls the keybd_event function. Two timers namely $T_o$ and $T_1$ are involved. $T_1$ is the interval which is the time required for acquiring a measurement from equipment/sensor. $T_o$ is the interval after which measurement is repeated. The virtual Keys sequence are sent to the main data acquisition program for acquiring data from measuring equipment, after which Timer $T_1$ is enabled. Upon completion of $T_1$ key sequence are sent to the main data acquisition program to save the file on the computer. The keyboard emulation software enters into a wait state equal to time interval $T_o$. After $T_o$ is elapsed then all the steps are repeated again. The flow chart representation of this algorithm is shown in Figure 7. In order to ensure that the destined data acquisition program window is selected win32 API FindWindow method is used which selects the appropriate window.

## V. APPLICATION AREAS

### A. Data Acquisitions at Hazardous or Non Accessible areas

Usage of the proposed system is best utilized for those areas where there is a requirement of data acquisition but human access is not possible. Example of such an environment include high electromagnetic environment. Moreover if there exists extreme weather conditions like snowy, windy or hot at the data acquisition place and prolonged logging of data is required then the system can be used efficiently. Radiated Immunity testing of systems in an anechoic chamber requires that systems should be operated in their normal operating modes also encompassing software operations. However, when electromagnetic field is radiated on the systems then physical user cannot operate the system. In such cases our proposed system can perform the software operations during the course of testing.

### B. Keyboard Emulator as part of Remote Data Acquisition System

Extensive research is being done on the use of remote instrumentation systems. In [11] numerous examples of implementation in this regard are given. Every device in a remote lab should be interoperable so that desired operations can be achieved by sending appropriate commands from remote locations. At the remote location a central computer controls the devices whenever a request from the remote user is received. If a device cannot be integrated with the main bus (GPIB, CAN etc) then our proposed system can provide the interoperability needs and provide the necessary interfacing through keyboard emulation. In the proposed grid based remote instrumentation system at [12], our designed system can be efficiently used with the remote instrumentation farm if any particular device cannot be readily integrated with the data acquisition bus. In this case our system can run the necessary steps by accepting the command string from the remote user at the destination end and perform the necessary operation through keyboard emulation.

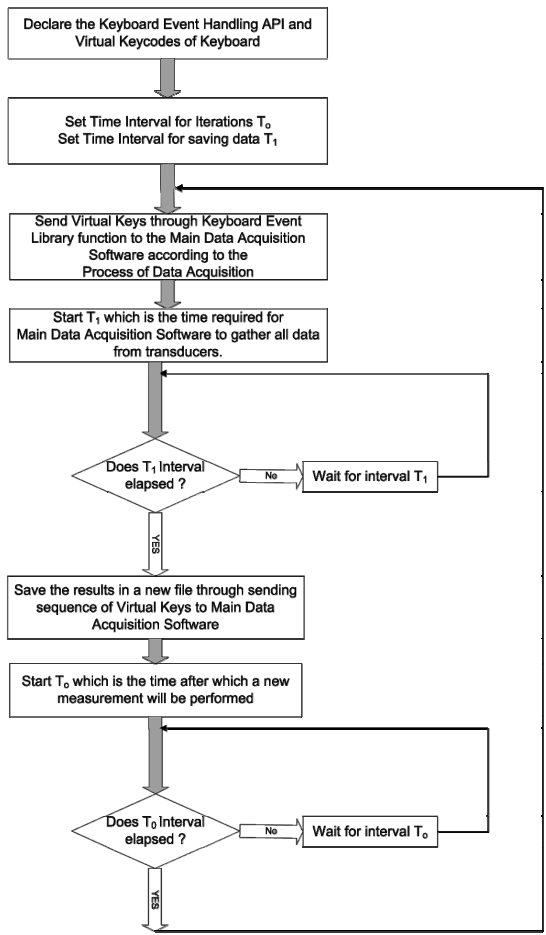

Figure 7: Flow chart of Keyboard Emulation software

## VI. CONCLUSION

Keybaord Emulation software can provide extended functionality to data acquisition process. The main data acquisition softwares can be operated as per end-user need by defining customizable Virtual Key sequence. This application can replace physical user with a virtual user which can run the data acquisition process through out its life cycle. Application areas of above implementation are but not limited to data acquisition at hazardous or non accessible places, solution of interoperability issues of devices and systems, helping disabled persons in using present day software applications etc. It is envisaged that the technique can be made efficient with the use of intelligent decision making technologies of neural networks and fuzzy logic.

REFERENCES

[1] Gabriel González, José P. Molina, Arturo S. García, Diego Martínez, and Pascual González, "Evaluation of Text Input Techniques in Immersive Virtual Environments", in *New Trends on Human–Computer Interaction*, Springer, 2009, pp 109-118.
[2] Ryszard Tadeusiewicz and Grażyna Demenko, "Speech Man-Machine Communication", in *Man-Machine Interaction*, Springer, 2009, pp 3-10.
[3] Darragh, John, and Ian Witten, "The reactive keyboard", Cambridge Univ Pr, 1992,pp 16-18.
[4] Mario Mühlehner, "Customizing User Interfaces with Input Profiles", in *Computers Helping People with Special Needs, Lecture Notes in Computer Science*, Springer, 2006. pp 442-449.
[5] Sporka, A.J., Kurniawan, S.H., Slavik, P, "Nonspeech Operated Emulation of Keyboard", in *Designing Accessible Technology*, Springer, 2006, pp 145-154.
[6] Altek Instruments Ltd(2010), *Altek Instruments - BarcodeMan - TheMule™*[Online],Available:http://www.barcodeman.com/altek/mule/.
[7] MicroRidge Systems Inc(2010), *Mircroridge Hardware and software measurements collections solution*[Online], Available: http://www.microridge.com/wedgelink.htm.
[8] Electracode(2010), *JoyToKey English Version*,[Online] ,Available:http://www.electracode.com/4/joy2key/JoyToKey%20English%20Version.htm.
[9] Microsoft Corporation(2010),*Keyboard Scan Code Specification Revision 1.3 a*,[Online], Available: http://download.microsoft.com/download/1/6/1/161ba512-40e2-4cc9-843a-923143f3456c/scancode.doc.
[10] Microsoft Corporation(2010), *Virtual Keycodes* [Online], Available:http://msdn.microsoft.com/en-us/library/ms927178.aspx
[11] F. Davoli et al. (eds.), in *Remote Instrumentation and Virtual Laboratories*, Springer, 2010.
[12] L. Berruti, F. Davoli, S. Vignola, and S. Zappatore, "Performance Analysis of a Grid-Based Instrumentation Device Farm, in *Remote Instrumentation and Virtual Laboratories,* Springer, 2010.